\documentclass[journal,letterpaper,twocolumn,final,10pt]{IEEEtran}

\usepackage{amsthm}
\usepackage{amsmath}
\usepackage{amsfonts}
\usepackage{dsfont}
\usepackage{mathrsfs}
\usepackage{pifont}
\usepackage{amssymb}
\usepackage{verbatim}
\usepackage{upgreek}
\usepackage{color}
\usepackage[final]{graphicx}
\usepackage{algorithmic}
\usepackage{algorithm}
\usepackage{epsfig}
\usepackage{eucal}
\usepackage{cite}
\usepackage{balance}

\newcommand{\bbm}{\begin{bmatrix}}
\newcommand{\ebm}{\end{bmatrix}}

\newcommand{\etr}[1]{{\mathrm{etr}}\left\{#1\right\}}

\newcommand{\bit}{\begin{itemize}}
\newcommand{\eit}{\end{itemize}}

\newcommand{\ben}{\begin{enumerate}}
\newcommand{\een}{\end{enumerate}}

\newcommand{\bdesc}{\begin{description}}
\newcommand{\edesc}{\end{description}}

\newcommand{\bea}{\begin{array}}
\newcommand{\eea}{\end{array}}

\newcommand{\beqa}{\begin{eqnarray}}
\newcommand{\eeqa}{\end{eqnarray}}

\newcommand{\ds}{\displaystyle}

\newcommand{\Comment}[1]{}

\def\Real{{\mathfrak Re}}
\def\C{{\mathds C}}

\def\cC{\mbox{$\CMcal C$}}

\def\cN{\mbox{$\CMcal N$}}

\newcommand{\be}{\begin{equation}}
\newcommand{\ee}{\end{equation}}

\newcommand{\bzero}{{\mbox{\boldmath $0$}}}

\newcommand{\bn}{{\mbox{\boldmath $n$}}}
\newcommand{\bm}{{\mbox{\boldmath $m$}}}

\newcommand{\bor}{{\mbox{\boldmath $r$}}}

\newcommand{\bv}{{\mbox{\boldmath $v$}}}
\newcommand{\bx}{{\mbox{\boldmath $x$}}}

\newcommand{\bz}{{\mbox{\boldmath $z$}}}

\newcommand{\bI}{{\mbox{\boldmath $I$}}}

\newcommand{\bM}{{\mbox{\boldmath $M$}}}

\newcommand{\bR}{{\mbox{\boldmath $R$}}}
\newcommand{\bS}{{\mbox{\boldmath $S$}}}

\newcommand{\bZ}{{\mbox{\boldmath $Z$}}}

\newcommand{\balpha}{{\mbox{\boldmath $\alpha$}}}

\newcommand{\test}{\mbox{$
\begin{array}{c}
\stackrel{ \stackrel{\textstyle H_{\hat{l},\hat{h}}}{\textstyle >} }{
\stackrel{\textstyle <}{\textstyle H_{0}} }
\end{array}
$}}

\title{Adaptive Radar Detection of Dim Moving Targets in Presence of Range Migration}

\author{Pia Addabbo,~\IEEEmembership{Member,~IEEE}, Danilo Orlando,~\IEEEmembership{Senior Member,~IEEE},  and Giuseppe Ricci,~\IEEEmembership{Senior Member,~IEEE}
\thanks{Pia Addabbo is with Universit\`a degli Studi ``Giustino Fortunato", viale Raffale Delcogliano, 12, 82100 Benevento, Italy. E-mail: {\tt p.addabbo@unifortunato.eu} }
\thanks{Danilo Orlando is with Universit\`a degli Studi ``Niccol\`o Cusano'', 
        Via Don Carlo Gnocchi, 3,  00166 Roma, Italy.
        E-mail: {\tt danilo.orlando@unicusano.it}}
\thanks{Giuseppe Ricci is with the Dipartimento di Ingegneria dell'Innovazione,
        Universit\`a del Salento, Via Monteroni, 73100 Lecce, Italy.
        E-mail: {\tt giuseppe.ricci@unisalento.it}.
        }
}

\begin{document}

\normalsize

\vspace{2cm}

\maketitle

\begin{abstract}
This paper addresses adaptive radar detection of dim moving targets. To circumvent range migration, the detection problem is formulated as 
a multiple hypothesis test and solved applying model order selection rules which allow to estimate the ``position'' 
of the target within the CPI and eventually detect it. 
The performance analysis proves
the effectiveness of the proposed approach also in comparison to existing alternatives.
\end{abstract}



\section{Introduction}
\label{sec:Introduction}

Adaptive radar detection of dim targets is a challenging problem in the radar community \cite{Richards}. 
It
plays a role of primary concern when target is under tracking by the radar. 
In this case, system resources
are suitably allocated for the tracking function and, from a tactical point of view, missing the target
would waste resource allocation time \cite{Richards}. Thus, it is desirable to 
collect as much energy as possible to increase the signal-to-interference-plus-noise ratio (SINR).
Otherwise stated, the ultimate performance depends on the number of processed pulses and, eventually, 
on the duration of the coherent processing interval (CPI); 
however, for moving targets the number of integrated pulses is limited by the range migration phenomenon.
For high resolution radars, which 
can resolve a target into a number of different scattering centers depending on the radar bandwidth and 
the range extent of the target \cite{ScheerMelvin}, and/or high-speed targets, 
range migration is an even more critical issue. 
Techniques to circumvent range migration are borrowed from synthetic aperture radar processing where  
the Keystone transform (KT) is used for moving-targets imaging \cite{5978178,5740946,4063317}. 
In fact, KT has also been applied to compensate range cell migration in the context of radar detection. 
In particular, for radar with low pulse repetition frequency (PRF) and, hence, in the presence 
of possible velocity ambiguities, \cite{KT_radar2005} applies a correction that depends 
on the folding factor of the target Doppler. 
A Keystone-like transform has also been proposed to perform coherent integration in \cite{Bidon2011}.
In \cite{EKT_radar2014}, the KT has been extended to compensate nonlinear range cell migration 
caused by radial acceleration of a maneuvering target. Further examples can be found 
in \cite{7855584,7112638,8115293}.

In this paper, we develop innovative architectures capable of detecting dim moving targets
resorting to model order selection (MOS) rules \cite{Stoica1}.
Specifically, we assume that the CPI has an adequate duration from the energy point of view. Since a 
moving target might enter and/or exit the cell under test (CUT), we formulate the detection problem as 
a multiple hypothesis test and, then, apply the MOS rules to estimate the position and the extension
of the target within the CPI. These estimates are then used to accomplish the detection task which
is either delegated to an additional stage or jointly performed along with estimation in a 
single-stage architecture. The performance analysis is conducted on simulated data and highlights
the advantage of the proposed architectures over the classical approaches for extended targets.

The paper is organized as follows:
the next section is
devoted to the problem formulation; Section \ref{Sec:Designs} contains the derivation 
of the detectors, whereas
Section \ref{Sec:Assessment} provides some numerical examples (also in comparison to natural competitors).
Concluding remarks are given in Section \ref{Sec:Conclusions}.

\subsection{Notation}

In the sequel, vectors and matrices are denoted by boldface lower-case and upper-case letters, respectively.
The symbols $|\cdot|$, $\det(\cdot)$, $\etr \cdot$, $(\cdot)^T$, $(\cdot)^\dag$ denote modulus value, determinant, 
exponential of the trace, transpose, and conjugate transpose, respectively. 
$\C$ is the set of complex numbers and $\C^{N\times M}$ is the Euclidean space of $(N\times M)$-dimensional complex matrices. The symbols $\Real\left\{ z \right\}$ indicates the real part of the complex  number $z$, 
$\bzero$ is the null vector of proper dimension, and 
$\bI_N$ stands for the $N \times N$ identity matrix.
Finally, we write $\bx\sim\cC\cN_N(\bm, \bM)$ if $\bx$ is an $N$-dimensional complex normal vector with mean $\bm$ and positive definite covariance matrix $\bM$.

\section{Problem formulation}
\label{sec_prob}
In this section, we first introduce the discrete-time signal  model
for a typical space-time scenario and then we formulate the detection problem as a multiple
hypothesis test. Note that the model is aimed at
taking into account range migration at the detection stage and it is obtained by properly modifying derivations in \cite{Bidon2011,BOR-Morgan}. The interested reader is referred to 
\cite{Ward-TR,Klemm} for an in-depth description of space-time adaptive processing.


Let us assume that the system  is equipped with a linear array of $N_a$
uniformly spaced and identical sensors deployed along the $z$
axis of a given reference system. Suppose that the $m$th sensor is 
located at $z_m=(m-1)d$, $m=1, \ldots, N_a$, with $d=\lambda/2$ and $\lambda$ denoting, in turn,  the operating
wavelength. 
The radar transmits a coherent burst of
$N_p$ radiofrequency (RF) pulses at a
constant $\mbox{PRF}=1/T$,
where $T$ is the pulse repetition time (PRT). Finally, the carrier
frequency is $f_c=c/\lambda$ where $c$
is the velocity of propagation in the
medium.
Due to the superposition principle,
we can leave aside for the moment the interference components and focus on
the useful target signal.
To this end, we suppose that the signal backscattered from the target is a delayed and attenuated copy of
the transmitted one. Specifically, suppose that the array is steered along a given direction, say $\psi$, 
measured with respect to the array direction, then
the signal transmitted along $\psi$ over the time interval $ [0, N_p T)$
is given by
\be
s^{tx}(t) =
\Real\left\{A e^{j\varphi}
\sum_{n=0}^{N_p-1}p(t^{\prime}_n)
e^{j2\pi f_c nT}
e^{j2\pi f_c t^{\prime}_n}\right\}
\ee
where $t^{\prime}_n=t-nT \in (0,T)$ is the fast time,
$A>0$ is an amplitude factor related to the transmitted power, 
$\varphi \in[0,2\pi)$ is a phase component depending on the local oscillator,
finally, $p(t)$ is a rectangular pulse of duration $T_p$ ($T_p$ is
much smaller than $T$).
If the radar  illuminates a
point-like target moving with constant radial velocity $v$ (with
$v>0$ for a target approaching the radar), the response to the $n$th pulse 
emitted by a sensor located at the origin of the reference system is the delayed version of the transmitted one  
by\footnote{We are neglecting the target displacement over the pulse duration.}
$\tau_n(t^{\prime}_n)= \tau_n - \frac{2v}{c}t^{\prime}_n$,
where 
$\tau_n= \tau_0 - \frac{2v}{c}nT$, with $\tau_0$, in turn, the round trip delay corresponding to the range at $t=0$, say $R_0$.

Thus, the received target echo at the
$m$th sensor is given by
\begin{eqnarray}
\nonumber
s^{rx}_{m}(t) &=&
\Real\Bigg\{\alpha \sum_{n=0}^{N_p-1}p\left(t^{\prime}_n-\tau_0 + \frac{2v}{c}nT+\frac{2v}{c}t^{\prime}_n \right)
\\ &\times & 
e^{j2\pi f_c nT}
e^{j2\pi
f_c[t^{\prime}_n-\tau_0 + \frac{2v}{c}nT+\frac{2v}{c}t^{\prime}_n+\Delta_m]} 
\Bigg\}
\label{eqn:ricevuto1}
\end{eqnarray}
where 
$\Delta_m=(m-1)d \cos \psi /c$
is the
travel time  between the $m$th
sensor and the origin while
$\alpha \in \C$ is a factor which accounts for $A e^{j\varphi}$, transmitting antenna gain, radiation pattern of the array sensors,
two-way
path loss, and radar cross-section of the (slowly-fluctuating)
target; hereafter, constant terms are absorbed into $\alpha$.

Neglecting the time scale compression or stretching of the transmitted pulses \cite{BOR-Morgan},
the target signal at the $m$th antenna element can also be written as
\begin{eqnarray}
s^{rx}_{m}(t) 
&=& 
\Real\Bigg\{ \alpha
\sum_{n=0}^{N_p-1}p\left(
t-nT
-\tau_0 + \frac{2 v}{c} nT\right) \nonumber
\\
&\times & 
e^{j2\pi (f_c+f_d)t}
e^{j 2\pi (m-1) \nu_s} \Bigg\}
\label{eqn:ricevuto3}
\end{eqnarray}
where $f_d=f_c \frac{2 v}{c}$
is the Doppler frequency shift of the possible target backscattered signal
and
$\nu_s$ is the target {\em spatial frequency}, given by $\nu_s=\frac{d}{\lambda} \cos{\psi}$.

As a consequence, after complex baseband conversion, the target
signal at the $m$th antenna element is given by
$$
x_{m}(t) = \alpha \!
\sum_{n=0}^{N_p-1} \!
p\left(
t
-nT -\tau_0 +\frac{2v}{c} nT 
\right) 
e^{j2\pi f_d t}
e^{j 2\pi (m-1) \nu_s}.
$$

A discrete form for the received signal
at the $m$th sensor is obtained by sampling the
output of a filter matched\footnote{We assume that the pulse waveform is Doppler tolerant.}
to $p(t)$ and fed by $x_{m}(t)$.
In particular, the matched filter output for the $m$th sensor is 
given by (recall that $p(\cdot)$ is a real pulse)
\begin{align}
y_m(t) &= x_m(t) * p(-t)= \alpha  e^{j 2\pi (m-1) \nu_s} 
\sum_{n=0}^{N_p-1} 
\int_{-\infty}^{+\infty}
 e^{j2\pi f_d u}
\nonumber
\\ 
&\times  p\left(
u-nT \left(1-\frac{2v}{c}\right)
-\tau_0\right) 
p(u-t)  du.
\end{align}
Letting $u_1=u-nT\left(1-\frac{2v}{c}\right)-\tau_0$, then
\begin{align*}
&y_m(t) = \alpha  e^{j 2\pi (m-1) \nu_s} 
\sum_{n=0}^{N_p-1} 
\int_{-\infty}^{+\infty}
e^{j2\pi f_d \left( u_1+nT\left(1-\frac{2v}{c}\right)+\tau_0\right)}
\\ 
&\times  p\left(
u_1\right) 
p\left(u_1-\left(t - nT\left(1-\frac{2v}{c}\right)-\tau_0 \right) \right) du_1
\\ 
&\approx  \alpha  e^{j 2\pi (m-1) \nu_s} \!\!\!
\sum_{n=0}^{N_p-1} \!\!\!
e^{j2\pi f_d T n}
\chi_p\left(t - nT\left(1-\frac{2v}{c}\right)-\tau_0,f_d\right)
\end{align*}
where $\chi_p(\cdot,\cdot)$ is the {\em ambiguity function} of the
pulse waveform $p(t)$~\cite{Levanon}.

In order to generate the range gate
corresponding to a round-trip delay $\tau_0+kT_p$,
$y_m(t)$ is sampled at the time instants
$t_{ki}=\tau_0+kT_p+iT$, $i=0,\dots,N_p-1$; 
we obtain
\begin{align}
y_m(t_{ki})
&= \alpha  e^{j 2\pi (m-1) \nu_s} 
\sum_{n=0}^{N_p-1} 
e^{j2\pi f_d T n} \nonumber
\\ 
&\times
\chi_p\left(kT_p+(i-n)T + nT\frac{2v}{c},f_d\right).
\end{align}
Note that $y_m(t_{ki})$ is nonzero only if  $k,i,n$ are such that
$
-T_p \leq kT_p+(i-n)T + nT\frac{2v}{c}  \leq T_p.
$
It follows that we have a nonzero sample for $n=i$ and $k=0$ if $|nT\frac{2v}{c}| \leq T_p$.
However, we typically
choose 
$k=0$ if $|nT\frac{2v}{c}| \leq T_p/2$,
$k=1$ if $-T_p \leq nT\frac{2v}{c} \leq -T_p/2$,
and $k=-1$ if $T_p/2 \leq nT\frac{2v}{c} \leq T_p$.
Moreover, we choose $k=1$ if $-3T_p/2 \leq nT\frac{2v}{c} \leq -T_p/2$
and $k=-1$ if $T_p/2 \leq nT\frac{2v}{c} \leq 3T_p/2$.
More generally, we choose $k=\overline{k}$ if
$-(2\overline{k}+1)T_p/2 \leq nT\frac{2v}{c} \leq -(2\overline{k}-1)T_p/2$.
Summarizing, given initial range
and velocity $v$ of the target, we are able to construct the corresponding time steering vector. 

From a different prospective, suppose that we want to test the possible presence of a target 
with unknown radial velocity $v$ within the CUT. Then, we can construct the 
following multiple-hypothesis testing problem
\be
\begin{cases}
H_{l,h} : &
\begin{cases}
\bz_1=\bn_1, \ldots, \bz_{l-1}=\bn_{l-1}
\\
\bz_l=\alpha_l \bv + \bn_l,\dots,\bz_{l+h}=\alpha_{l+h} \bv + \bn_{l+h}
\\
\bz_{l+h+1}=\bn_{l+h+1}, \ldots, \bz_{N_p}=\bn_{N_p}
\\
\bor_k=\bm_k, \ k=1,\ldots,K
\end{cases}
\\
H_0 : &
\begin{cases}
\bz_1=\bn_1, \ldots, \bz_{N_p}=\bn_{N_p}
\\
\bor_k=\bm_k, \ k=1,\ldots,K
\end{cases}
\end{cases}
\label{eqn:detectionProblem}
\ee
where 
$\bv\in\C^{N_a\times 1}$ is the spatial steering vector while
$l\in\Omega=\{1,\ldots,N_p\}$ 
and $h\in\{0,\ldots,N_p-1\}$, $l+h\leq N_p$, are integers indexing those spatial vectors containing target components\footnote{Note that the target might enter and/or exit the CUT within the dwell time.}. In fact, the $\bn_i$s and the $\bm_j$s are noise vectors that we model as independent random vectors; moreover, we suppose\footnote{Otherwise stated, the $\bor_k$ are training vectors that we assume homogeneous to those from the CUT.} that
$\bn_i,\bm_j \sim\cC\cN_{N_a}(\bzero,\bM)$.
{We also assume that $K \geq N_a$} and that the $\alpha_i\in\C$, $i=l,\ldots,l+h$, are unknown (deterministic) complex factors.

Two remarks are in order. First,
observe that the above problem subsumes a noncoherent data integration whose efficiency depends
on the correlation degree amid the interference returns. Moreover, the latter is tied up to
several factors as the PRF, the aspect angle, the radar frequency agility, and the dwell time \cite{Richards}.
Finally, it is important to underline that problem \eqref{eqn:detectionProblem} contains 
several possibly nested alternate hypotheses.
In the next section, we exploit MOS rules to devise two classes of adaptive architectures 
for problem \eqref{eqn:detectionProblem}.

\section{Detector Designs}
\label{Sec:Designs}
The herein proposed architectures differ in the number of stages. Specifically, the first architecture 
consists of a preliminary stage which provides
estimates for $l$ and $h$, followed by a second stage, devoted to the detection, which exploits the above
estimates to form
a suitable decision statistic. 
The second architecture jointly performs detection and estimation by incorporating the objective function
of the considered MOS rule into a sort of generalized likelihood ratio test (GLRT) based decision statistic.

For both architectures, we choose the generalized information 
criterion (GIC) that provides a tuning parameter
allowing for a decrease of the overfitting probability and can overcome the Akaike information criterion \cite{Stoica1}. 
In addition, we discard the Bayesian information criterion since it would lead to a 
prior-dependent rule, which has little practical value \cite{Stoica1}.

Before proceeding with the design, for future reference, let us introduce some useful definitions. Specifically, let
$\bZ=[\bz_1 \cdots \bz_{N_p}]$, $\bR=[\bor_1 \cdots\bor_K]$, $\Omega_{l,h}=\{l,\ldots,l+h\}$ and denote by
\begin{multline*}
f_{l,h}(\bZ;\balpha,\bM)=[\pi^{N_aN_p} \det(\bM)^{N_p}]^{-1}
\\
\times
\ds \mbox{etr}\left\{ -\bM^{-1}
\left[\sum_{i\in\Omega_{l,h}} (\bz_i-\alpha_i\bv)(\bz_i-\alpha_i\bv)^\dag + \bS_{l,h}^{\prime} \right] \right\}
\end{multline*}
with $\bS_{l,h}^{\prime}=\sum_{i\in\Omega\setminus\Omega_{l,h}}\bz_i\bz_i^\dag$ and $\balpha=[\alpha_l \cdots \alpha_{l+h}]^T$,
$f_{0}(\bZ;\bM)={ \mbox{etr}\{ -\bM^{-1}
[\bZ\bZ^\dag] \}}/
[\pi^{N_aN_p} \det(\bM)^{N_p}]$,
and
$
f(\bR;\bM)=
{\ds \mbox{etr}\{ -\bM^{-1}
[\bR\bR^\dag] \}}/
[\pi^{N_aK} \det(\bM)^{K}]$ 
the probability density function (PDF) of $\bZ$ under $H_{l,h}$, the PDF of $\bZ$ under $H_0$, and the PDF
of $\bR$, respectively.

\subsection{Two-stage Architectures}
As previously stated, in this class of architectures, the first stage estimates $l$ and $h$, whereas the second stage is responsible for the detection task. 
Two approaches are followed in building up the selection rule. The first approach consists in deriving 
the GIC rule for known $\bM$, which is then replaced with the sample covariance based upon the training vectors. 
For this reason we refer to this rule as two-step GIC.
The second approach computes the maximum likelihood estimates of $\balpha$ and $\bM$ using the joint PDF of $\bZ$ and $\bR$.
Thus, according to the first approach, the expression of GIC for known $\bM$ is given by
\be
\min_{l=1,\ldots,N_p  \atop h: \ l+h\leq N_p} 
\left\{-\log f_{l,h}(\bZ;\widehat{\balpha}(\bM),\bM) + p_1(h,\rho)\right\}
\label{eqn:GIC00}
\ee
where $p_1(h,\rho)=(1+\rho)(h + 1)$, with $\rho > 1$ the tuning parameter and
$$
\widehat{\balpha}(\bM)=\left[\frac{\bv^\dag\bM^{-1}\bz_l}{\bv^\dag\bM^{-1}\bv} \cdots
\frac{\bv^\dag\bM^{-1}\bz_{l+h}}{\bv^\dag\bM^{-1}\bv}\right]^T
$$
the maximum likelihood estimate (MLE) of $\balpha$ for known $\bM$. Replacing $\bM$ with $\bS/K=\bR\bR^\dag/K$,
it is possible to show that (\ref{eqn:GIC00}) is equivalent to the following minimization problem
\begin{align}
&\min_{l=1,\ldots,N_p \atop h: \ l+h\leq N_p} 
-K\sum_{i\in\Omega_{l,h}} \frac{|\bz_i^\dag\bS^{-1}\bv|^2}
{\bv^\dag\bS^{-1}\bv} + p_1(h,\rho).
\label{eqn:GIC00_A}
\end{align}
In the second case, GIC rule becomes
\be
\min_{l=1,\ldots,N_p \atop h: \ l+h\leq N_p} \!\!\!\!\!
\left\{-2\log [f(\bR;\widehat{\bM}_{l,h})
f_{l,h}(\bZ;\widehat{\balpha}_{l,h},\widehat{\bM}_{l,h})] + p_2(h,\rho)\right\}
\label{eqn:GIC01}
\ee
where $p_2(h,\rho)=(1+\rho)[2(h+1)+N_a^2]$,
$$\widehat{\bM}_{l,h}= \frac{\bS_{l,h}
+\sum_{i\in\Omega_{l,h}}(\bz_i-\widehat{\balpha}_{l,h}(i)\bv)(\bz_i-\widehat{\balpha}_{l,h}(i)\bv)^\dag}{N_p+K}
$$
is the MLE of $\bM$,
based upon $\bZ$ and $\bR$, under $H_{l,h}$,
$\bS_{l,h}=\bR\bR^\dag+\sum_{i\in\Omega\setminus
\Omega_{l,h}}\bz_i\bz_i^\dag$, 
and $\widehat{\balpha}_{l,h}(i)=
\frac{\bv^\dag\bS_{l,h}^{-1}\bz_i}{\bv^\dag\bS_{l,h}^{-1}\bv}$
is the $i$th entry of $\widehat{\balpha}_{l,h}$, the MLE of $\balpha$, using $[\bZ \ \bR]$,
under $H_{l,h}$. Finally, it is possible to show that \eqref{eqn:GIC01} is
equivalent to
\be
\min_{l=1,\ldots,N_p \atop h: \ l+h\leq N_p} 
\left\{2(N_p+K)\log \det(\widehat{\bM}_{l,h}) + p_2(h,\rho)\right\}.
\label{eqn:GIC02}
\ee
Once an estimate of $(l,h)$, say $(\hat{l},\hat{h})$, is available, then the second stage compares a 
statistic which is function of $(\hat{l},\hat{h})$ with a detection threshold. Specifically, we consider the
following decision statistic
\be
\sum_{i=\hat{l}}^{\hat{l}+\hat{h}} \frac{|\bz_i^\dag\bS^{-1}\bv|^2}
{\bv^\dag\bS^{-1}\bv} \test \eta
\ee
where here and after $\eta$ is the threshold set to ensure the desired value for the probability 
of false alarm ($P_{fa}$).

\subsection{One-stage Architectures}
The estimation stage developed in the previous subsection can be suitably modified in 
order to provide it with detection capabilities 
making the second stage unnecessary. To this end, we exploit a GLRT like approach where the PDF under the alternate
hypothesis is replaced by the MOS objective function, namely a penalized compressed likelihood, due 
to the fact that in this case there exist multiple alternate hypotheses. 

The first architecture is derived proceeding exactly as for the two-step GIC \eqref{eqn:GIC00_A}
and leads to the detector
\be
\max_{l=1,\ldots,N_p \atop h: \ l+h\leq N_p} 
K \sum_{i\in\Omega_{l,h}} \frac{|\bz_i^\dag\bS^{-1}\bv|^2}
{\bv^\dag\bS^{-1}\bv} - p_1(h,\rho)\test \eta.
\ee
The second detector is obtained by considering GIC based upon $\bZ$ and $\bR$. Specifically,
it is given by
\begin{multline}
\max_{l=1,\ldots,N_p \atop h: \ l+h\leq N_p} 
\left\{\log [f(\bR;\widehat{\bM}_{l,h})
f_{l,h}(\bZ;\widehat{\balpha}_{l,h},\widehat{\bM}_{l,h})] - \frac{p_2(h,\rho)}{2}\right\}
\\
-\max_{\bM}\log[f(\bR;\bM)f_{0}(\bZ;\bM)]\test\eta
\end{multline}
and is equivalent to
\begin{multline}
\log\det((\bS+\bZ\bZ^\dag)/(N_p+K))
\\
+\max_{l=1,\ldots,N_p \atop h: \ l+h\leq N_p} 
\left\{-\log \det(\widehat{\bM}_{l,h}) - \frac{p_2(h,\rho)}{2(N_p+K)}\right\}\test\eta.
\end{multline}

\section{Performance assessment}
\label{Sec:Assessment}

In this section, we investigate the behavior of both one- and two-stage architectures through
numerical examples. The considered performance metrics are the probability of detection ($P_d$) and the root mean square
error (RMSE) in the estimation of $l$ and $h$. For simulations purposes,
we resort to standard Monte Carlo counting techniques by evaluating 
the thresholds to ensure $P_{fa}=10^{-4}$, the $P_d$s, and RMSE values over $100/P_{fa}$, $1000$, and $1000$ independent 
trials, respectively.  
At each trial, the values of $h$ and $l$ are uniformly generated in $[0,N_{p}-1]$ and $[1,N_p-h]$, respectively. 
The interference covariance matrix is given by $\bM = \sigma^2_n\bI_{N_a} +  p_c \bM_{c}$, where $\sigma^2_n\bI_{N_a}$ represents the thermal noise component with power $\sigma^2_n=1$, 
while  $p_c \bM_{c}$ is the clutter component 
with $p_c$ the clutter power (set 
assuming a clutter-to-noise ratio of 20 dB) and $\bM_c$ the clutter covariance matrix, whose $(i,j)$th entry 
is given by $\bM_c(i,j)=\rho_c^{|i-j|}$ with $\rho_c=0.95$. 
Finally, the SINR is defined as $\textrm{SINR}=|\alpha|^2\bv^{\mathrm{\dag}}\bM^{-1}\bv$.

\begin{figure}
    \centering
    \includegraphics[width=.38\textwidth]{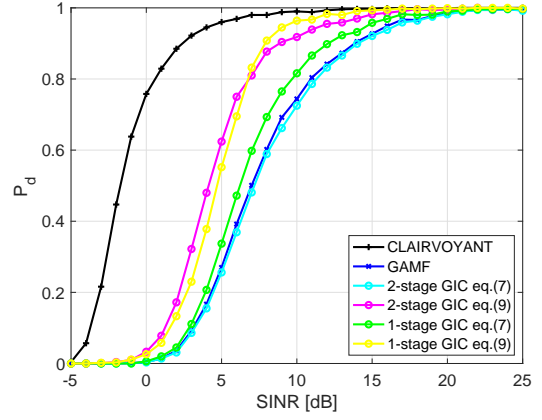}
    \caption{$P_d$ versus SINR for $N_p=16$, $N_a=8$, and $K=16$.}
    \label{fig:pd}
\end{figure}

\begin{figure}
    \centering
    \includegraphics[width=.38\textwidth]{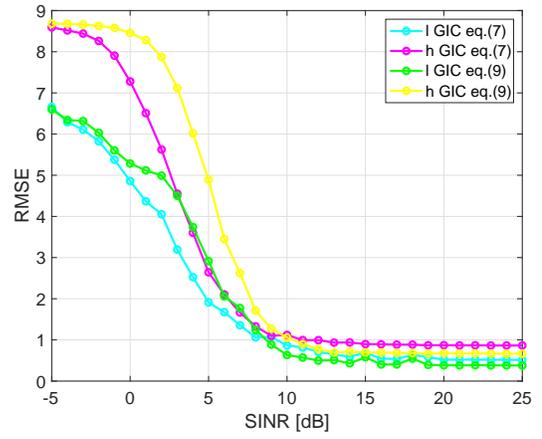}
    \caption{RMSE versus SINR for $N_p=16$, $N_a=8$, and $K=16$.}
    \label{fig:rmse}
\end{figure}

The GIC tuning parameters in \eqref{eqn:GIC00} and \eqref{eqn:GIC01}
are equal to 11 and 5, respectively\footnote{These values represent a 
reasonable compromise to limit the model overestimation for both cases.}.
For comparison purposes, we have also plotted the $P_d$ curves of the so-called generalized adaptive matched filter 
(GAMF) introduced in \cite{conte01} and the clairvoyant (non-adaptive) detector for known $\bM$, $l$, and $h$, 
which represents an upperbound to the detection performance. Fig. \ref{fig:pd} shows that architectures
based upon \eqref{eqn:GIC01} ensure better detection performance with a gain of about 3 dB over the other
decision schemes. On the contrary, the GAMF along with the 2-stage architecture based on \eqref{eqn:GIC00}
are placed in the last position of the performance rank.
In Figure \ref{fig:rmse}, we show that for SINR values greater than $10$ dB the estimation error of the considered
GIC-based architectures is less than 1.


\section{Conclusions}
\label{Sec:Conclusions}
In this letter, we have addressed that problem of detection in the presence of range migration. To this end,
we have devised adaptive architectures which incorporate MOS rules to estimate the position and 
the extension of the target within the CPI.
The performance analysis has highlighted the effectiveness of the proposed approach, since decision schemes
based upon \eqref{eqn:GIC01} can provide a significant performance gain with respect to the GAMF 
and the other herein proposed detectors. From the estimation point of view, the considered architectures can
ensure a negligible RMSE for SINR values greater than $10$ dB. Finally, future research tracks might encompass
the design of tracking algorithms capable of accounting for target range migration between different scans.

\balance
\bibliographystyle{IEEEbib}
\bibliography{biblio}

\begin{thebibliography}{10}

\bibitem{Richards}
M.~A. Richards, J.~A. Scheer, and W.~A. Holm,
\newblock {\em {Principles of Modern Radar: Basic Principles}},
\newblock Raleigh, NC, 2010.

\bibitem{ScheerMelvin}
W.~L. Melvin and J.~A. Scheer,
\newblock {\em {Principles of Modern Radar: Advanced Techniques}},
\newblock Edison, NJ, 2013.

\bibitem{5978178}
C.~{Dai}, X.~{Zhang}, and J.~{Shi},
\newblock ``{Range Cell Migration Correction for Bistatic SAR Image
  Formation},''
\newblock {\em IEEE Geoscience and Remote Sensing Letters}, vol. 9, no. 1, pp.
  124--128, Jan. 2012.

\bibitem{5740946}
Y.~{Jungang}, H.~{Xiaotao}, J.~{Tian}, J.~{Thompson}, and Z.~{Zhimin},
\newblock ``{New Approach for SAR Imaging of Ground Moving Targets Based on a
  Keystone Transform},''
\newblock {\em IEEE Geoscience and Remote Sensing Letters}, vol. 8, no. 4, pp.
  829--833, Jul. 2011.

\bibitem{4063317}
D.~{Zhu}, Y.~{Li}, and Z.~{Zhu},
\newblock ``{A Keystone Transform Without Interpolation for SAR Ground
  Moving-Target Imaging},''
\newblock {\em IEEE Geoscience and Remote Sensing Letters}, vol. 4, no. 1, pp.
  18--22, Jan. 2007.

\bibitem{KT_radar2005}
S.~Zhang, T.~Zeng, T.~Long, and H.~Yuan,
\newblock ``Dim target detection based on keystone transform,''
\newblock in {\em IEEE International Radar Conference}, May 2005, pp. 889--894.

\bibitem{Bidon2011}
S.~{Bidon}, L.~{Savy}, and F.~{Deudon},
\newblock ``Fast coherent integration for migrating targets with velocity
  ambiguity,''
\newblock in {\em 2011 IEEE Radar Conference}, May 2011, pp. 027--032.

\bibitem{EKT_radar2014}
X.~{Tian}, S.~{Zhang}, and L.~{Pang},
\newblock ``Range cell migration correction for dim maneuvering target
  detection,''
\newblock in {\em 2014 IEEE Radar Conference}, May 2014, pp. 1247--1250.

\bibitem{7855584}
X.~{Li}, G.~{Cui}, W.~{Yi}, and L.~{Kong},
\newblock ``{Fast coherent integration for maneuvering target with high-order
  range migration via TRT-SKT-LVD},''
\newblock {\em IEEE Trans. on Aerospace and Electronic Systems}, vol. 52, no.
  6, pp. 2803--2814, Dec. 2016.

\bibitem{7112638}
L.~{Kong}, X.~{Li}, G.~{Cui}, W.~{Yi}, and Y.~{Yang},
\newblock ``{Coherent Integration Algorithm for a Maneuvering Target With
  High-Order Range Migration},''
\newblock {\em IEEE Trans. on Signal Processing}, vol. 63, no. 17, pp.
  4474--4486, Sept. 2015.

\bibitem{8115293}
F.~{Pignol}, F.~{Colone}, and T.~{Martelli},
\newblock ``{Lagrange-Polynomial-Interpolation-Based Keystone Transform for a
  Passive Radar},''
\newblock {\em IEEE Trans. on Aerospace and Electronic Systems}, vol. 54, no.
  3, pp. 1151--1167, Jun. 2018.

\bibitem{Stoica1}
P.~Stoica and Y.~Selen,
\newblock ``{Model-order selection: A review of information criterion rules},''
\newblock {\em IEEE Signal Processing Magazine}, vol. 21, no. 4, pp. 36--47,
  Jul. 2004.

\bibitem{BOR-Morgan}
F.~{Bandiera}, D.~{Orlando}, and G.~{Ricci},
\newblock {\em {Advanced Radar Detection Schemes Under Mismatched Signal
  Models}},
\newblock {Morgan and Claypool Publishers}, 2009.

\bibitem{Ward-TR}
J.~Ward,
\newblock ``Space-time adaptive processing for airborne radar,''
\newblock Tech. {R}ep. 1015, Lincoln Laboratory, MIT, Lexington, MA, Tech.
  Rep., Dec. 1994.

\bibitem{Klemm}
R.~Klemm,
\newblock {\em {Principles of Space-Time Adaptive Processing}},
\newblock IEE Radar, Sonar, Navigation and Avionics Series 12, 2002.

\bibitem{Levanon}
N.~Levanon,
\newblock {\em {Radar Principles}},
\newblock John Wiley and Sons, 1988.

\bibitem{conte01}
E.~{Conte}, A.~{De Maio}, and G.~{Ricci},
\newblock ``{GLRT-based adaptive detection algorithms for range-spread
  targets},''
\newblock {\em IEEE Trans. on Signal Processing}, vol. 49, no. 7, pp.
  1336--1348, Jul. 2001.

\end{thebibliography}

%

\end{document}